\documentclass[twocolumn,superscriptaddress,pra]{revtex4}

\usepackage{amsmath,amssymb,amsfonts,bm}
\usepackage{graphicx,epstopdf}  
\usepackage{color}
\usepackage{lscape}
\usepackage{extarrows}
\usepackage{enumerate}
\usepackage{empheq}
\usepackage{multirow}

\usepackage{array}
\usepackage{booktabs}

\newcommand{\M}{\mbox{\tiny M}}

\newcommand{\w}{\omega}
\newcommand{\wti}{\widetilde}
\newcommand{\ti}{\tilde}

\newcommand{\B}{\mbox{\tiny B}}

\newcommand{\tS}{\mbox{\tiny S}}

\newcommand{\T}{\mbox{\tiny T}}

\newcommand{\env}{\mbox{\tiny env}}

\newcommand{\sol}{\mbox{\tiny sol}}

\newcommand{\la}{\langle}
\newcommand{\ra}{\rangle}

\newcommand{\Sec}[1]{Sec.\,\ref{#1}}

\newcommand{\nl}{\nonumber \\}
\newcommand{\be}{\begin{equation}}
\newcommand{\ee}{\end{equation}}
\newcommand{\bsube}{\begin{subequations}}
\newcommand{\esube}{\end{subequations}}
\newcommand{\Eq}[1]{Eq.\,(\ref{#1})}
\newcommand{\Eqs}[1]{Eqs.\,(\ref{#1})}
\newcommand{\Fig}[1]{Fig.\,\ref{#1}}

\newcommand{\RN}[1]{%
  \textup{\uppercase\expandafter{\romannumeral#1}}%
}

\begin{document}
\title{
Extended dissipaton theory for higher-order bath couplings and application to non-Condon spectroscopy with anharmonicity
}

\author{Zi-Fan Zhu}
\affiliation{Hefei National Research Center for Physical Sciences at the Microscale and Department of Chemical Physics, University of Science and Technology of China, Hefei, Anhui 230026, China}

\author{Yu Su}
\affiliation{Hefei National Research Center for Physical Sciences at the Microscale and Department of Chemical Physics, University of Science and Technology of China, Hefei, Anhui 230026, China}

\author{Yao Wang}
\email{wy2010@ustc.edu.cn}
\affiliation{Hefei National Research Center for Physical Sciences at the Microscale and Department of Chemical Physics, University of Science and Technology of China, Hefei, Anhui 230026, China}

\author{Rui-Xue Xu}
\email{rxxu@ustc.edu.cn}
\affiliation{Hefei National Research Center for Physical Sciences at the Microscale and Department of Chemical Physics, University of Science and Technology of China, Hefei, Anhui 230026, China}

\author{YiJing Yan}
\affiliation{Hefei National Research Center for Physical Sciences at the Microscale and Department of Chemical Physics, University of Science and Technology of China, Hefei, Anhui 230026, China}

\date{\today}
\begin{abstract}
In this work, we develop an extended dissipaton theory that generalizes the environmental couplings beyond the conventional linear and quadratic forms, enabling the treatment of arbitrary order of bath couplings. Applying this theoretical framework to the condensed-phase non-Condon spectroscopy, we demonstrate the interplay of anharmonicity, non-Condon and solvent effects on optical spectra. Precise simulations are carried out with high efficiency on linear absorption spectra involving the above mentioned correlated effects. We exhibit how an anharmonic potential modulates the vibronic feature, offering insights into the role of nonlinear environmental couplings in spectroscopic signatures and exemplifying the success of the extended dissipaton formalism as an exact and efficient method for higher-order bath couplings.

\end{abstract}
\maketitle

\section{Introduction}
\label{sec1}

Anharmonicity is inevitable in molecules.
For example, highly excited anharmonic oscillators, observed via gas-phase high-resolution infrared spectroscopy, may behave as local modes \cite{Zhu88181,Zhu89596,Zhu902691,Chi9141, Chi93116,Che971079}. 
One may be interested in the related effect on the non-Condon electronic spectroscopy, for example the Herzberg–Teller type of excitation, in condensed phase. 
Long--lived quantum beats have been observed in photosynthetic antenna complexes \cite{Eng07782,Pan1012766}.
It is also found that solvent and
non-Condon effects may enhance
quantum coherence \cite{Kam0010637,Ish082219,Che21244105}.

Consider the anharmonic effect on
non-Condon spectroscopy in solvents. The total Hamiltonian under the interaction with external field may read
\begin{align}\label{Htot}
 H_{\T}(t) = H_{\M}-\varepsilon(t)\hat \mu_{\T},
\end{align}
where
\begin{align}\label{HMtot}
 H_{\M} =
 H_{\env}|g\ra\la g|+(\Delta+H'_{\env})|e\ra\la e|.
\end{align}
Here, $\Delta$ is the excitation energy. The total electronic--plus--vibrational dipole moment $\hat \mu_{\T}$ interacts with a classical external field $\varepsilon(t)$. $H_{\env}$ and $H'_{\env}$ are the vibration--plus--solvent environment Hamiltonians,  associated with the ground $|g\ra$ and excited $|e\ra$ states, respectively.
To apply methods
of dissipative systems, one need recast the total matter Hamiltonian into the
system--plus--environment  form as 
\be\label{HT0}
  H_{\M} = \Delta|e\ra\la e| 
  +(H'_{\env}-H_{\env})|e\ra\la e| +H_{\env}\,.
\ee
%
In this work,
we will consider the potential in $H'_{\env}$ to be of anharmonicity,
which will lead to the coupling between the electronic--state system and the 
vibration--plus--solvent environment, resulted from 
$H'_{\env}-H_{\env}$, be non-linear beyond quadratic hence non-Gaussian.

However, most quantum dissipation theories describe just the reduced system under the influence of Gaussian environment in the regime of linear coupling.
Therefore, the theoretical study here encounters the challenge that the system is coupled to a 
non-Gaussian environment which also interacts with the external field. Furthermore, the excitation of environment will interfere with the system.
One may think to include vibrational modes into the system together with the electronic states. This apparently increase the dimension of simulation and computing consumption,
especially with the anharmonicity and the influence of solvent, to obtain accurate results.
In this work we will compromise
the strategy that the vibrational mode and solvent are treated consistently and compatibly as the role of environment (thermal bath).

We have developed an extended dissipaton-equation-of-motion (DEOM) approach \cite{Xu18114103,Zhu25234103} for non-linear bath couplings but so far only 
explicitly established up to the quadratic order.
In this work, we will emphasize on non-Gaussian, anharmonic bath effect, which goes beyond the quadratic order. We will present the generalized DEOM formalism for arbitrary order of bath couplings.
The DEOM 
\cite{Yan14054105,Zha15024112}
adopts quasi-particle descriptions for environments,
which provides a unified and exact treatment on not only the system dissipative evolutions but also the collective environment dynamics
and entangled system--environment excitations.
The DEOM method has been applied to study the correlated vibration--solvent effects on the non-Condon spectroscopy 
in the Gaussian, linear coupling bath domain previously \cite{Che21244105}, where the coherence enhancement due to the synergetic vibration–solvent correlation was clearly demonstrated.
For the focus of this paper on non-Gaussian, anharmonic bath and also for clarity, we will concentrate on the Brownian--vibration condition in 
Ref.\onlinecite{Che21244105},
by which the vibrational mode in the solvent behaves as a Brownian
oscillator while the solvent does not directly affect the electronic transition.

This paper is organized as follows.
The general DEOM treatment
on arbitrary nonlinear bath couplings beyond the quadratic order
are elaborated in \Sec{thsec2}.
Numerical demonstrations are presented in \Sec{thsec3} on non-Condon spectra.
The paper is summarized in \Sec{thsec4}.
Throughout this paper, we set the Planck constant and Boltzmann constant as units ($\hbar=1$ and $k_B=1$), and $\beta=1/T$, with $T$ being the temperature.

\section{The DEOM approach for arbitrary-order bath couplings}
\label{thsec2}
This section presents a general theoretical framework based on the dissipaton algebra  to deal with higher-order environmental couplings. 
In this work, these higher-order couplings arise from the anharmonicity of nuclear potential surface of the excited state.
Under the Brownian--vibration condition \cite{Che21244105},
\begin{align} 
H'_{\env}-H_{\env}= V'(\hat q')-V(\hat q)
\end{align}
where
\begin{align}\label{eq_potential}
V(\hat{q})=\frac{\Omega}{2} \hat{q}^2\ \ \ \ \text{and}\ \  \ \ V'(\hat{q}')&=\frac{\Omega}{2} \hat{q}^{\prime 2}+\sum_{r\geq 3}\alpha_r \hat{q}^{\prime r},
\end{align}
with $\hat q'=\hat q-D$.
Explicitly,
\begin{align}
V'(\hat q')-V(\hat q)&=\bigg[\frac{\Omega}{2} (\hat q-D)^2+\sum_{r\geq 3}\alpha_r (\hat q-D)^r\bigg]-\frac{\Omega}{2} \hat q^2
\nl &
\equiv \sum_{r\geq 0} \bar\alpha_r \hat q^r
\end{align}
where
\begin{align}
\bar\alpha_0=&\frac{\Omega D^2}{2}+\sum_{r\geq 3}\alpha_r (-D)^r,
\\
\bar \alpha_1=&-\Omega D+\sum_{r\geq 3}\alpha_r r (-D)^{r-1},
\end{align}
and
\begin{align}
\bar \alpha_{s}=\sum_{r\geq {\rm max}(3,s)}\alpha_r {r\choose s} (-D)^{r-s}\ \ \ \ (s\geq 2).
\end{align}
Here, we set $\hat q$ to be dimensionless, and all $\{\alpha_r\}$ and $\{\bar \alpha_r\}$ have the dimensions of energy.
Under the Brownian--vibration condition \cite{Che21244105}, the bath  spectral density reads
\be\label{chiqqw}
   \wti\chi_{qq}(\w)
    =\frac{\Omega}{\Omega^2-\w^2-i\w\ti\zeta(\w)},
\ee
where $\ti\zeta(\w)$ is the solvent induced frictional function. 

The correlation function can be obtained via the fluctuation-dissipation theorem,
\be\label{FDT}
\la \hat q(t)\hat q(0)\ra_{\B}=\frac{1}{\pi}\int_{-\infty}^{\infty}\!\!{\rm d}\w\,\frac{{\rm Im}\wti \chi_{qq}(\w)}{1-e^{-\beta \w}}e^{-i\w t}.
\ee
The DEOM construction starts with an exponential expansion of correlation function satisfying \Eq{FDT} as
\be\label{FF_exp0}
\la \hat q(t)\hat q(0)\ra_{\B}
 = \sum_{k} \eta_{k} e^{-\gamma_{k} t}.
\ee
This can generally be achieved via certain
sum--over--poles decomposition \cite{Hu10101106} on the Fourier transform function
in \Eq{FDT}, followed by the Cauchy's contour
integration in the low--half plane, or the time-domain Prony fitting schemes \cite{Che22221102}.
The resulting $\{\gamma_k\}$
are either real
or complex--conjugate paired. We can then
 define the associated index $\bar k$ via
$\gamma_{\bar k}\equiv \gamma^{\ast}_{k}$.
Due to the time--reversal relation,
$\la \hat q(0)\hat q(t)\ra_{\B}=\la \hat q(t)\hat q(0)\ra_{\B}^{*}$,
we have
\be\label{FF_exp_rev0}
\la \hat q(0)\hat q(t)\ra_{\B}
= \sum_{k} \eta^{\ast}_{{\bar k}}e^{-\gamma_{k} t} .
\ee

The dissipaton decomposition
on the environment operator reads \cite{Yan14054105}
\be\label{F_in_f}
 \hat q = \sum_{k}\hat f_{k}.
\ee
This decomposition recovers \Eqs{FF_exp0} and (\ref{FF_exp_rev0}), by assuming that dissipatons are statistically independent,
with their correlation functions $(t>0)$
\be\label{ff_corr}
\begin{split}
 \la \hat f_{k}(t)\hat f_{k'}(0)\ra_{\B}
 &= \delta_{kk'}\eta_{k}e^{-\gamma_{k} t},
\\
 \la \hat f_{k'}(0)\hat f_{k}(t)\ra_{\B}
 &= \delta_{kk'}\eta^{\ast}_{{\bar k}}e^{-\gamma_{k} t}.
\end{split}
\ee
 The dynamical variables in DEOM are called the dissipaton density operators (DDOs),
defined as
\be\label{DDO}
 \rho^{(n)}_{\textbf{n}}(t)\equiv {\rm tr}_{\B}\Big[
  \Big(\prod_{k} \hat f^{n_{k}}_{k}\Big)^\circ
  \rho_{\T}(t)\Big].
\ee
Here, $n=\sum_{k} n_{k}$ and $\textbf{n}=\{n_{k}\}$
that is an ordered set of the occupation numbers, $n_{k}=0,1,\cdots$,
on individual dissipatons.
The circled parentheses, $(\cdots)^{\circ}$, is the irreducible notation.
For bosonic dissipatons it follows that
$(\hat f_{k}\hat f_{k'})^{\circ}=(\hat f_{k'}\hat f_{k})^{\circ}$.
The key ingredient in the dissipaton algebra
is the generalized Wick's theorem:
\bsube\label{Wick12}
\begin{align}\label{Wick1}
 &\quad \rho_{\bf n}(t;f_{k'}^{>}) \equiv  {\rm tr}_{\B}\Big[\Big(\prod_{k} \hat f^{n_{k}}_{k}\Big)^\circ
   \hat f_{k'} \rho_{\T}(t)\Big]
\nl&=
 \rho^{(n+1)}_{{\bf n}^{+}_{k'}}(t)+\sum_{k} n_{k}\la\hat f_{k}\hat f_{k'}\ra^{>}_{\B}
   \rho^{(n-1)}_{{\bf n}^{-}_{k}}(t),
\end{align}
and
\begin{align}\label{Wick2}
 &\quad \rho_{\bf n}(t;f_{k'}^{<}) \equiv {\rm tr}_{\env}\Big[\Big(\prod_{k} \hat f^{n_{k}}_{k}\Big)^\circ
   \rho_{\T}(t)\hat f_{k'} \Big]
\nl&=
 \rho^{(n+1)}_{{\bf n}^{+}_{k'}}(t)+\sum_{k} n_{k}\la\hat f_{k'}\hat f_{k}\ra^{<}_{\B}
   \rho^{(n-1)}_{{\bf n}^{-}_{k}}(t).
\end{align}
\esube
Here, ${\bf n}^{\pm}_{k}$ differs from ${\bf n}$ only
at the specified $\hat f_{k}$-disspaton occupation number,
$n_{k}$, by $\pm 1$
and
\be\label{ff0}
\begin{split}
 \la\hat f_{k}\hat f_{k'}\ra^{>}_{\B}
 &\equiv \la\hat f_{k}(0+)\hat f_{k'}(0)\ra_{\B} = \eta_{k}\delta_{kk'},
\\
 \la\hat f_{k'}\hat f_{k}\ra^{<}_{\B}
 &\equiv \la\hat f_{k'}(0)\hat f_{k}(0+)\ra_{\B} = \eta^{\ast}_{{\bar k}}\delta_{kk'}.
\end{split}
\ee
 The DEOM can then be readily constructed
by applying the Liouville equation, $\dot{\rho}_{\T}(t)=-i[H_{\T}(t),\rho_{\T}(t)]$,
to the total composite density operator in \Eq{DDO};
i.e.,
\be\label{DDO_dot}
 \dot\rho^{(n)}_{\textbf{n}}(t)= -i\, {\rm tr}_{\B}\Big\{
  \Big(\prod_{k} \hat f^{n_{k}}_{k}\Big)^\circ
  [H_{\T}(t),\rho_{\T}(t)]\Big\}.
\ee

For the problem studied in this work, we assume the Herzberg-Teller type of non-Condon coupling between the composite matter and the external field [cf.\,\Eq{Htot}], i.e.,
\be \label{dipole} 
\hat \mu_{\T}=\hat\mu_{\tS}\otimes (1+v_{\B}\hat q),
\ee
where $\hat\mu_{\tS}=u_{\tS} (|g\ra \la e|+|e\ra\la g|)$ is the system dipolar operator and $v_{\B}$ accounts for the non-Condon contribution.
We then recast the total Hamiltonian,
\Eq{Htot}, as
\begin{align}\label{HT}
  & \quad\ H_{\T}(t) =H_{\tS}(t)+H_{\env}
+\hat Q\sum_{r\geq 1} \bar\alpha_r \hat q^r-v_{\B}\hat \mu_{\tS}\hat q\varepsilon(t)
\nl &
= H_{\tS}(t)\!+\!H_{\env}
\!+\wti Q(t)(\sum_k \hat f_k)
\!+\hat Q\sum_{r\geq 2} \bar\alpha_r  (\sum_k \hat f_k)^r,
\end{align}
where
\be 
H_{\tS}(t)= \Delta|e\ra\la e|-\hat \mu_{\tS}\varepsilon(t)+\bar\alpha_0\hat Q,
\ee
with
$
\hat Q=|e\ra \la e|
$,
and 
\be 
\wti Q(t)=\bar \alpha_1\hat Q-v_{\B}\hat \mu_{\tS}\varepsilon(t).
\ee

We then construct from \Eq{DDO_dot} the final formalism of DEOM reading
\begin{align}\label{DEOM}
 \dot\rho^{(n)}_{\bf n}(t)&=-i[H_{\tS}(t),\rho^{(n)}_{\bf n}(t)]-i\rho^{(n)}_{\bf n}(t;H_{\env}^{\times})
 \nl & \quad
 -i\sum_k\Big[\wti Q(t) \rho^{(n)}_{\bf n}(t;\hat f_k^{>})-\rho^{(n)}_{\bf n}(t;\hat f_k^{<})\wti Q(t) \Big]
  \nl &  \quad 
 -i\sum_{r\geq 2}\sum_{{\bf r}|r}\bar\alpha_r \Big[\hat Q \rho^{(n)}_{\bf n}\Big(t;\prod_k (\hat f_k^{>})^{r_k}\Big)
 \nl & \quad \quad \ 
 -\rho^{(n)}_{\bf n}\Big(t;\prod_k (\hat f_k^{<})^{r_k}\Big)\hat Q \Big],
\end{align}
where ${\bf r}=\{r_k\}$ with $r_k=0,1,2\cdots$, and $\sum_{{\bf r}|r}$ means that the sum is over all configurations of ${\bf r}$ under the constraint $\sum_{k}r_k=r$.
The second term on the r.h.s. of \Eq{DEOM} is evaluated via 
 the generalized diffusion equation \cite{Yan14054105},
\be\label{gendiff}
 i\rho^{(n)}_{\bf n}(t;H_{\env}^{\times})
=\Big(\sum_{ k} n_{k} \gamma_{ k}\Big)
  \rho^{(n)}_{\bf n}(t),
\ee
which is essentially proved via the phase-space dissipaton algebra \cite{Wan20041102}. 
The third and fourth terms can be elaborated and programmed by using the generalized Wick's theorem, \Eqs{Wick12}--(\ref{ff0}), iteratively. We thus have finished the construction of extended DEOM with arbitrary higher-order environmental couplings.

\section{Numerical demonstrations}\label{thsec3}

\begin{figure}[t]
\includegraphics[width=\columnwidth]{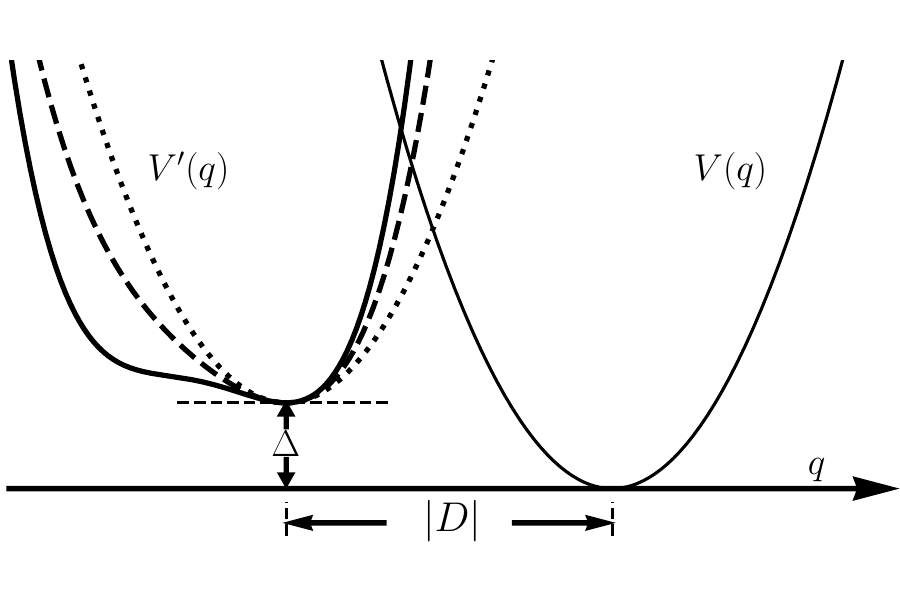}
\caption{Ground and excited potential energy surfaces. The dotted curve represents the harmonic reference ($g=0$) of the excited state. The dashed and solid curves exhibit the anharmonic excited-state potential surfaces with 
$g=1$ and 
$g=2$, respectively.
}\label{fig1}
\end{figure}

\begin{figure}[t]
\includegraphics[width=\columnwidth]{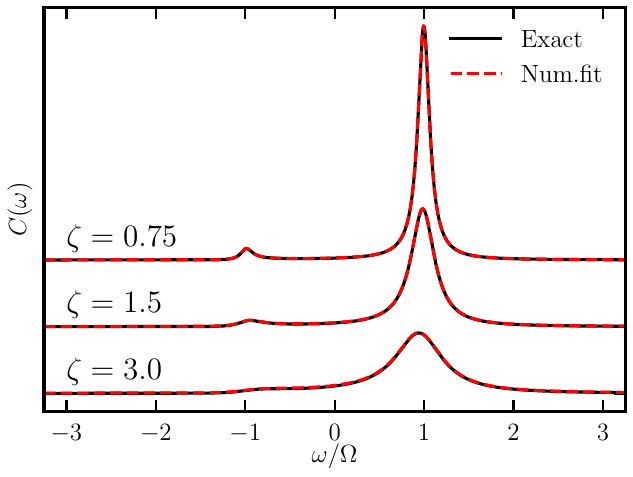}
\caption{The numerically fitted  bath spectra $C(\w)$ compared with exact ones, the Fourier transform of $\la \hat q(t)\hat q(0)\ra_{\B}$, with three specified values
of $\zeta=3\Omega$, $1.5\Omega$, $0.75\Omega$.
}\label{fig2}
\end{figure}

For numerical demonstrations, we choose the anharmonic potential function from Ref.\,\onlinecite{Hoe1879} as the reference for the  excited state.
In \Eq{eq_potential}, we select
$\Omega=632$cm$^{-1}$, $\alpha_3=g\times 52$cm$^{-1}$, $\alpha_4=g\times5.2$cm$^{-1}$, and $\alpha_{r> 4}=0$.
To exhibit the effect of anharmoncity, we vary the coefficient $g=0, 0.5,1,1.5, 2$ in simulations.
We plot the ground and excited potential surfaces in \Fig{fig1}. 
To show the non-Condon effect, we select $v_{\B}=0,0.5,1,1.5,2$ in \Eq{dipole}.
The temperature is chosen as 
$T = 298$K. In \Eq{chiqqw}, we assume the white-noise limit for the solvent influence, i.e., $\ti \zeta(\w)\approx \ti\zeta(0)\equiv \zeta$. Simulations are carried out with $\zeta=3\Omega$, $1.5\Omega$, $0.75\Omega$, and compared to the gas-phase results. The decomposition in \Eq{FF_exp0} adopts the numerical least-squares fit for the bi-exponential bath correlation expansion, as described in Ref.~\onlinecite{Din16204110}. The fitting accuracy is shown in \Fig{fig2}, where the fitted results are found to match the exact ones perfectly. The following simulating results are converged with the hierarchy tier of DEOM.
The simulating time is within a few seconds on an ordinary personal computer.

\begin{figure}[!t]
\includegraphics[width=0.9\columnwidth]{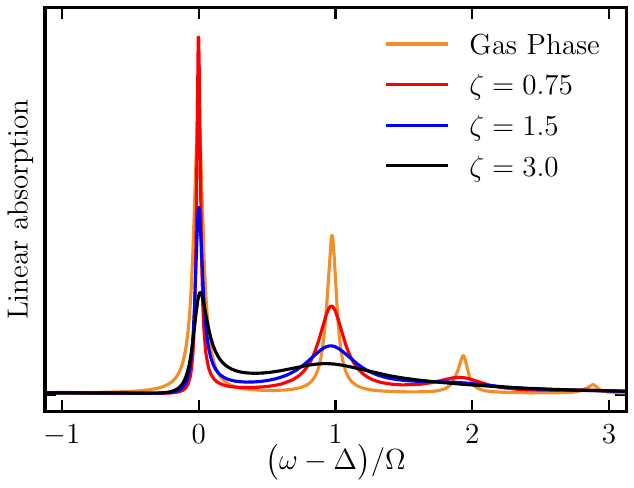}
\caption{The evaluatedlinear  absorption spectra 
with varied $\zeta$ and fixed $v_{\B}=g=1$. The gas phase result is computed via the standard Fermi golden rule approach, where
the $\delta$-functions have been broadened as Lorentzian functions.}\label{fig3}
\end{figure}

\begin{figure}[!t]
\includegraphics[width=0.9\columnwidth]{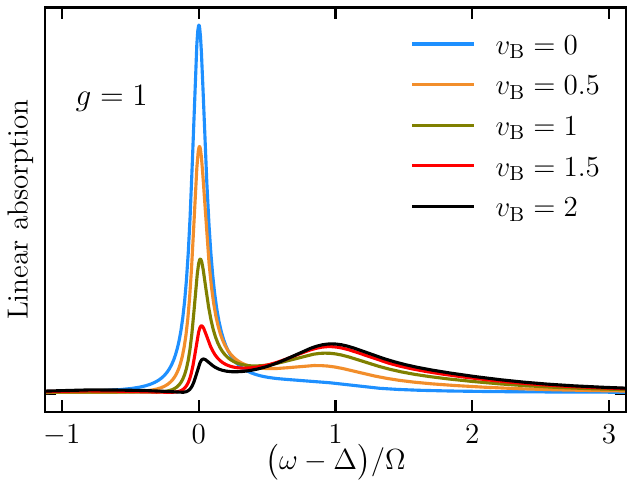}
\caption{The evaluated linear  absorption spectra 
with varied $v_{\B}$ and fixed $g=1$.}\label{fig4}
\end{figure}

\begin{figure}[!t]
\includegraphics[width=0.9\columnwidth]{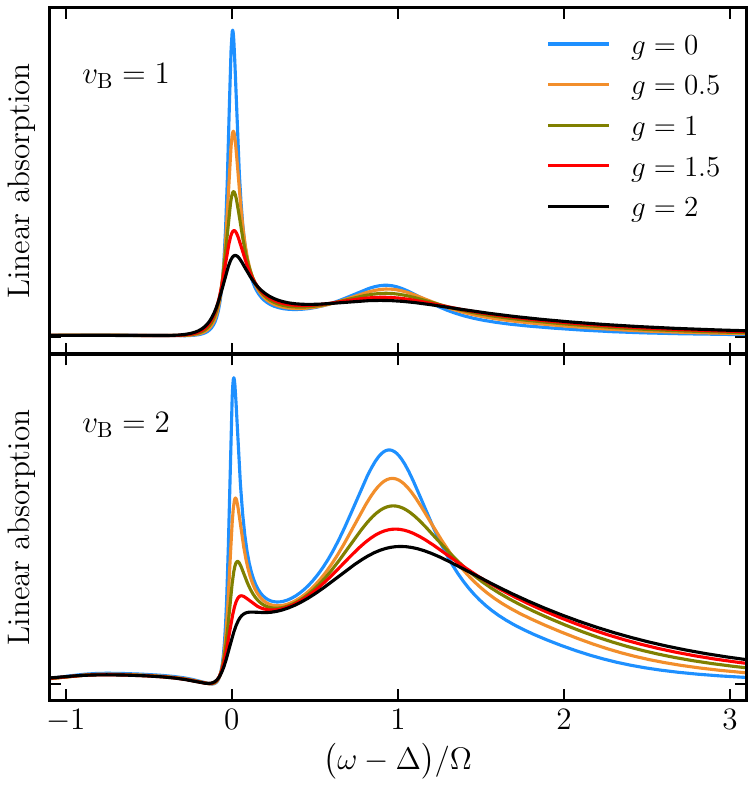}
\caption{The evaluated linear  absorption spectra 
with varied $g$ and fixed $v_{\B}=1$ (upper panel) and $2$ (lower panel).}\label{fig5}
\end{figure}

Figure \ref{fig3} depicts the evaluated linear  absorption spectra 
with fixed $v_{\B}=g=1$ and varied $\zeta$, to exhibit the solvent broadening effect. The gas phase result is computed via the standard Fermi golden rule approach, where
the $\delta$-functions have been broadened as Lorentzian functions. All the spectra shown in the figure are scaled to be of the same integration area. 
The results shown in the following \Fig{fig4} and \Fig{fig5} will be calculated with $\zeta=3\Omega$.

Figure \ref{fig4} shows the calculated linear absorption spectra for various values of $v_{\B} $ with a fixed anharmonicity coefficient of $ g = 1$.  As illustrated, the non-Condon effect enhances the relative intensity of the vibrational peak (around $1$) compared to the electronic excitation peak (around $0$). 
Figure~\ref{fig5} presents the linear absorption spectra for different values of $g$ with fixed $ v_{\B} = 1$ (upper-panel) and $2$ (lower-panel). As the anharmonicity increases, both the electronic excitation peak near $0$ and the vibrational peak near $1$ are gradually broadened and less symmetrized.

\section{Summary}\label{thsec4}

In this work, we have developed an extended dissipaton theory which systematically generalizes the treatment of bath couplings beyond the conventional linear or quadratic forms. By including arbitrary orders of bath couplings, the theoretical framework presented in this paper establishes a unified and exact approach to describe more complex bath-induced effects such as anharmonicity which were previously inaccessible by quantum dissipation methods. This theoretical advancement provides thus a versatile platform for exploring various forms of  system–bath interactions and their implications in diverse quantum dynamical processes.

Numerically the method has been successfully applied to the exact simulation of the condensed-phase non-Condon spectroscopy involving anharmonicity.
The simulation is precise and carried out with a high computing efficiency. The interplay between the anharmonicity of the vibrational mode,
the solvent and the non-Condon effects is demonstrated.
The present extended dissipaton theory for arbitrary bath couplings is shown to be promising and feasible for the study of open quantum systems embedded in complex
environments with anharmonicity, bath induced mode–mode correlation, non-Gaussian fluctuations, etc. These complex systems and environments are commonly encountered in biological pigments, molecular aggregates, and solid-state devices, etc. The integration of the present theory with \textit{ab initio} parameterizations and machine-learned potential energy surfaces would further enhance its theoretical applicability to realistic systems. 


\begin{acknowledgments}
Support from  the National Natural Science Foundation of China (Grant Nos.\   
22373091, 224B2305 \& 22573099) is gratefully acknowledged. Simulations were
performed on the robotic AI-Scientist platform of Chinese
Academy of Sciences.
\end{acknowledgments}

\appendix


\end{document}